\documentclass[10pt]{article}

\usepackage[utf8]{inputenc}	
\usepackage[T1]{fontenc}
\usepackage[left=1.45in,right=1.45in,bottom=1in,top=1in]{geometry} 
\usepackage{makeidx}         
\usepackage{graphicx}        
\usepackage{multicol}        
\usepackage[bottom]{footmisc}
\usepackage[colorlinks = true,
            linkcolor = blue,
            urlcolor  = blue,
            citecolor = blue,
            anchorcolor = blue]{hyperref}
\usepackage{amsmath,amssymb}

\usepackage{lipsum}
\usepackage{tikz}
\usepackage{pgfplots}
\usepackage{pgfplotstable}
\usepackage{booktabs,colortbl}
\usepackage{authblk}
\usepackage{xspace}
\usepackage{microtype}

\usetikzlibrary{shadows,calc,spy}
\newcommand\bs\boldsymbol
\newcommand\dblbrackl{\text{\textlbrackdbl}}
\newcommand\dblbrackr{\text{\textrbrackdbl}}
\definecolor{green}   {RGB}{ 87 171  39}
\definecolor{green-10}{RGB}{242 247 236}
\definecolor{orange}   {RGB}{246 168   0}
\definecolor{orange-10}{RGB}{255 247 234}
\definecolor{red}   {RGB}{204   7  30}
\definecolor{red-10}{RGB}{250 235 227}

\title{Marking strategies for adaptive mesh refinement: An efficiency-focused benchmark study for steady solid and fluid mechanics problems}

\author{Oliver Wege\thanks{\href{mailto:wege@cats.rwth-aachen.de}{\texttt{wege@cats.rwth-aachen.de}}}}
\author{Kaan Atak}
\author{Marek Behr}
\author{Norbert Hosters}

\affil{Chair for Computational Analysis of Technical Systems (CATS), RWTH Aachen University, Aachen, Germany}

\begin{document}
\maketitle
\begin{abstract}%
Adaptive mesh refinement (AMR) is indispensable for efficient finite element analyses. 
However, its performance depends not only on the refinement itself but also on strategy to mark elements for refinement and the way it is tuned.
This work compares classical marking methods (maximum, Dörfler bulk-chasing, quantile) with non-classical, statistically based approaches (z-score, Isolation Forest), all driven by the residual-based Kelly error estimator and tested on steady solid and fluid mechanics problems. 
The study finds quantile and z-score markings to be the most robust, Dörfler effective for large bulk parameters, and maximum marking sensitive to irregular fields. 
Isolation Forest can rival top classical methods with a high contamination level but may fail under aggressive settings. 
These results offer practical guidance for selecting marking strategies that balance refinement aggressiveness and computational cost in adaptive FEM workflows.
\end{abstract}
\section{Introduction}
\label{sec:introduction}
Adaptive mesh refinement (AMR) is an indispensable tool in\textemdash{}but certainly not only\textemdash{}the finite element method (FEM) for efficiently resolving localized features and improving the general approximation quality in solutions of partial differential equations (PDEs) \cite{ainsworth1997posteriori,verfurth2013posteriori}.
The effectiveness and efficiency of AMR hinges on two critical components~\cite{dorfler1996convergent,morin2002convergence}:
\begin{itemize}
  \item the \textit{error estimator}, which quantifies the local discretization error, and
  \item the \textit{marking strategy}, which selects elements for refinement based on this error
\end{itemize}
The latter one will be addressed in this work.
While residual-based error estimators such as the \textit{Kelly} estimator \cite{kelly1983a,deSrGago1983b} have become standard practice \cite{kronbichler2012high,bangerth2007deal}, the choice of the marking strategy is frequently guided by user experience or determined through time-consuming trial-and-error approaches, rather than systematic analysis. 
This has significant implications for computational efficiency and solution accuracy, particularly when the valuable information provided by the error estimator is not leveraged sufficiently in the refinement process \cite{stevenson2007optimality}.
Classical marking strategies include maximum marking, bulk-chasing (such as \textit{Dörfler}'s $\theta$-marking), and equidistribution approaches \cite{stevenson2007optimality,cascon2008quasi}. 
These methods have been rigorously analyzed for their convergence properties and optimality \cite{stevenson2007optimality,binev2004adaptive}, yet their practical performance can vary considerably across different problem classes and error distributions.
In addition, other non-classical marking techniques based on statistical or machine learning approaches have been proposed.
In statistical approaches, the standard deviation of error indicators is used as a threshold to mark elements for refinement or coarsening \cite{grave2021adaptive}. 
Another approach employs an unsupervised learning method, the Isolation Forest, for anomaly detection as a marking strategy \cite{falini2021adaptive}. 
Beyond marking, machine learning is applied to optimize AMR workflows, such as using reinforcement learning to determine marking strategy parameters \cite{gillette2024learning} or recurrent neural networks to replace both estimation and marking steps \cite{bohn2021recurrent}. 
However, systematic comparisons between classical and non-classical strategies, particularly those leveraging statistical outlier detection such as $z$-score-based marking and Isolation Forest algorithms \cite{liu2008isolation}, remain scarce.
In this work, we present a comprehensive numerical study of marking strategies for AMR using the finite element method, focusing on their effectiveness and efficiency for stationary problems in solid and fluid mechanics.
Using the residual-based \textit{Kelly} error estimator \cite{kelly1983a,deSrGago1983b} adapted for vector field problems, we compare classical strategies (maximum marking \cite{verfurth2013posteriori}, quantile-based and \textit{Dörfler} marking \cite{dorfler1996convergent}) with non-classical $z$-score and outlier detection methods \cite{liu2008isolation}.
Our benchmarks include well-known problems from both fields, enabling a cross-disciplinary evaluation.
The scope of our study encompasses several key performance aspects: 
\begin{itemize}
  \item the $L^2$-norm (fluids) and the strain energy norm (solids) error convergence, 
  \item statistical view on the element size and the error indicator field distribution,
  \item influence of the threshold parameters in each marking strategy, and,
  \item number of refinement cycles required to reach an error threshold of $1\%$.
\end{itemize}
This multifaceted evaluation provides insight into both the effectiveness and computational efficiency of the marking strategies under consideration.
By systematically analyzing these metrics, we aim to identify strengths and limitations of both classical and non-classical marking strategies, offering guidance for their application in real-world FEM simulations.
\section{Adaptive mesh refinement (AMR)}
\begin{figure}
  \centering
  \includegraphics[width=.99\textwidth]{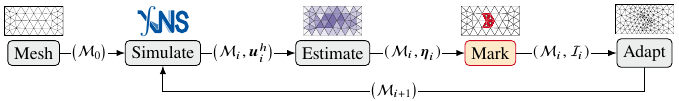}
  \caption{Iterative AMR: starting from an initial mesh $\mathcal{M}_0$, a simulation on $\mathcal{M}_i$ yields the approximated PDE solution $\bs{u}_i^h$, from which error indicators $\bs\eta_i$ are computed; a marking strategy selects regions with large error, i.e. a subset of element $\mathcal{I}_{i}$ for mesh adaptation. This process is then repeated until a desired convergence criterion is met.}
  \label{fig:amrworkflow} 
\end{figure}

A general AMR process is inherently iterative, consisting of repeated cycles of solution, error estimation, marking, and mesh adaptation as demonstrated in Fig. \ref{fig:amrworkflow}. 
At each refinement iteration $i$, the mesh $\mathcal{M}_i$ is used to compute an approximate solution $\bs{u}_i^h$, from which error indicators $\bs\eta_i$ are derived (Section \ref{ssec:refmetric}). 
A marking strategy then selects a subset of elements $\mathcal{I}_{i}\!\subseteq\!\mathcal{M}_{i}$ for refinement to adapt the mesh, aiming to reduce the discretization error (Section \ref{ssec:markingstrategies}).
There are two approaches to adapt the mesh in terms of mesh-to-mesh relations:
\begin{itemize}
  \item The mesh after refinement is a new mesh without a strong topological relation between the meshes \cite{verfurth2013posteriori}.
  \item The original mesh coexists with refined elements at multiple levels of resolution. 
        The finite element basis may be constructed as a sum of hierarchical basis functions on different levels \cite{bangerth2003adaptive,zander2015multi}.
\end{itemize}
Both have their own advantages and disadvantages, but we assume that both can utilize the same marking strategies with similar benefits. 
In this work, we will use the first approach driven by the meshing tool \texttt{Gmsh} \cite{geuzaine2009gmsh}.
The loop of refinement cycles is repeated until a desired convergence criterion is met which is usually based on the magnitude of a global vector space norm or the change of the solution between refinement cycles.
\subsection{Residual-based error estimator}
\label{ssec:refmetric}
In each refinement iteration $i$, the approximate solution $\bs{u}_i^h$ on mesh $\mathcal{M}_i$ is evaluated using an error estimator.
Specifically, the \textit{Kelly} estimator computes local error indicators $\eta_e$ for each mesh element $e$, based on the jumps in the solution gradient across element boundaries \cite{kelly1983a,deSrGago1983b}. 
These local indicators are collected in the error indicator vector $\bs\eta_i=\bs\eta(\mathcal{M}_i,\bs u_i^h)=[\eta_e]_i$, which is used to guide mesh adaptation (see Fig. \ref{fig:amrworkflow}).
The \textit{Kelly} error estimator was originally proposed for scalar field problems, but can be adapted in a straight-forward fashion for vector field problems which then suggests the element indicator
\begin{equation}
  \eta_e = \sqrt{\,
    \tfrac{\lvert\partial\Omega_e\rvert}{24}
    \int_{\partial\Omega_e}  \bigl\lVert\dblbrackl\nabla\bs{u}^h\dblbrackr\bigr\rVert_2^2 \,d\partial\Omega_e
  }
\end{equation}
wherein $\dblbrackl\nabla\bs{u}^h\dblbrackr$ is the jump of the approximated solution (displacement or velocity) gradient across the element boundary to its neighboring element due to the $\mathcal{C}^0$-continuity of standard finite element discretizations.
The pre-integral factor for 2D problems shown here\textemdash{}including the length of the element boundary $\lvert\partial\Omega_e\rvert$\textemdash{}is a common choice for scaling $\eta_e$ according to their element size.
\begin{center}
  \begin{minipage}{.9\textwidth}
    \textit{Remark:} For simplicity, we use $\bs{u}$ to denote the general primary unknown of a PDE (displacement or velocity), as the error estimator applies identically to both problems (solid and fluid mechanics).
  \end{minipage}
\end{center}
\subsection{Marking strategies}
\label{ssec:markingstrategies}
The element-wise error indicators $\eta_e$ form a spatial error distribution over the mesh. 
Based on these indicators, a marking strategy selects a subset of elements $\mathcal{I}_i\subseteq\mathcal{M}_i$ for refinement. 
A mesh is considered optimal when the local error indicators are approximately equidistributed across the mesh \cite{verfurth1994posteriori,nochetto2009theory}. 
The goal of a marking strategy is therefore to identify elements with high error values and refine them, promoting a more uniform error distribution. 
For efficiency, the number of marked elements should be kept as small as possible while maintaining effective error reduction \cite{nochetto2009theory}. 
Different marking strategies exist with different criteria for selecting the subset $\mathcal{I}$. 
In this study, we consider both classical marking strategies, such as maximum, quantile, and \textit{Dörfler} marking, and non-classical strategies, such as $z$-score and Isolation Forest.
\subsubsection{Maximum (\texttt{MAX})}
\label{sssec:maximum}
The maximum marking strategy, also known as the greedy marking strategy \cite{gillette2024learning}, defines the subset by selecting elements whose error indicators exceed a fixed fraction of the maximum error indicator value:
\begin{equation}
  \mathcal{I} = \left\{ e \in \mathcal{M}\,:\,
  \eta_e \ge \alpha \max\limits_{e^\prime\in\mathcal{M}} (\eta_{e^\prime}) \right\}
\end{equation}
\begin{center}
  \begin{minipage}{.9\textwidth}
    \centering
    \textit{``Mark all elements $e$ whose error indicator $\eta_e$ exceeds a fraction $\alpha\in[0,1]$ of the maximum error indicator in the mesh.''}
  \end{minipage}
\end{center}
In this way, the strategy targets regions with the highest error values. 
The parameter $\alpha \in [0,1]$ controls the aggressiveness of the refinement. With decreasing $\alpha$, more elements are selected. 
As $\alpha$ decreases to $0$, all elements are marked, while for $\alpha = 1$, only elements attaining the maximum indicator value are selected. 
By construction, this strategy always marks at least one element corresponding to the maximum error indicator \cite{morin2008basic}. 
Its performance depends on how dominant the maximum value is compared to the rest of the mesh.
\subsubsection{Quantile (\texttt{QUA})}
\label{sssec:quantile}
For the quantile marking strategy \cite{hennig2017adaptive}, the error indicator values are sorted in increasing order, and all elements above the $\gamma$-th quantile are selected for refinement:
\begin{equation}
  \mathcal{I} = \left\{ e \in \mathcal{M}\,:\,
  \eta_e \ge Q_\gamma \right\}
\end{equation}
\begin{center}
  \begin{minipage}{.9\textwidth}
    \centering
    \textit{``Mark all elements $e$ whose error indicator $\eta_e$ exceeds the $\gamma$-th quantile $Q_\gamma$ of all error indicators, with $\gamma\in[0,1]$.''}
  \end{minipage}
\end{center}
The strategy always selects a fixed fraction of elements to be refined, regardless of the absolute magnitude of the maximum error indicator. 
As a result, the method is less sensitive to extreme values and provides a robust selection of regions with high error. Similar to the maximum marking strategy, this approach always selects at least one element for refinement.
\subsubsection{Dörfler (\texttt{DOE})}
\label{sssec:doerfler}
The Dörfler marking strategy \cite{dorfler1996convergent} aims to reduce the total error by marking only the most significant contributors:
\begin{equation}
  \text{Find minimal }\mathcal{I} \subseteq \mathcal{M}\text{ such that }\sum\limits_{e\in\mathcal{I}} \eta_e^2 \ge \theta\sum\limits_{e^\prime\in\mathcal{M}} \eta_{e^\prime}^2
\end{equation}
\begin{center}
  \begin{minipage}{.9\textwidth}
    \centering
    \textit{``Mark a minimal subset $\mathcal{I} \subseteq \mathcal{M}$ of elements $e$ whose error indicator $\eta_e$ sum up to a fixed fraction $\theta \in [0,1]$ of the total square derror.''}
  \end{minipage}
\end{center}
Unlike the maximum or quantile strategies, which select elements based on individual error values, Dörfler's marking considers the cumulative contribution of elements to the total error.  
The elements are first sorted in descending order of their error indicators, and then the minimal subset is selected such that the sum of their squared errors reaches at least a fraction $\theta$ of the total squared error. 
Here, $\theta$ controls the aggressiveness of the refinement. 
A larger $\theta$ marks more elements to address a greater portion of the total error, while smaller $\theta$ focuses only on the largest contributors. 
By construction, this strategy guarantees that the selected elements account for most of the total error \cite{nochetto2009theory}. 
Nevertheless, the parameter selection $\theta$ remains crucial for efficiency. 
\subsubsection{Z-score (\texttt{ZSC})}
\label{sssec:zscore}
The $z$-score marking strategy \cite{grave2021adaptive} selects elements based on their deviation from the mean of the error indicators:
\begin{equation}
  \mathcal{I} = \left\{ e \in \mathcal{M}\,:\,
  z_e = \tfrac{\eta_e - \mathrm{mean}\eta_{e^\prime}}{\mathrm{stdev}\eta_{e^\prime}} \ge z^* \right\}
  \,,~~e^\prime\in\mathcal{M}
\end{equation}
\begin{center}
  \begin{minipage}{.9\textwidth}
    \centering
    \textit{``Mark elements $e$ whose error indicator $\eta_e$ has a $z$-score $z_e$ above a threshold $z^*$ ($z^*>0$ for upper and $z^*<0$ for lower outlier).''}
  \end{minipage}
\end{center}
It aims to identify outliers and generate equidistributed error over the mesh. 
The error indicators are first transformed using a logarithmic scaling to better capture relative variations. 
Elements are marked if their corresponding $z$-score exceeds a prescribed upper threshold $z^*$, i.e., if their error deviates from their mean error value by more than $z^*$ standard deviations.
The parameter $z^*$ controls the aggressiveness of the refinement. 
The smaller values lead to more elements being selected, whereas the larger values restrict marking to only the most outlying ones. 
Compared to classical strategies, like maximum and Dörfler, this approach is independent of the values of the error indicators and instead relies on their statistical distribution. 
Its effectiveness depends on the underlying distribution and for large threshold values, it may fail to mark any elements.
\subsubsection{Isolation Forest (\texttt{ISO})}
\label{sssec:isoforest}
The Isolation Forest \cite{liu2008isolation} is an unsupervised machine learning method for detecting anomalies, i.e., outliers, in a dataset.
By recursively partitioning the data, the outliers can be identified since they require fewer splits to be isolated.
The error indicators are treated as input data to the Isolation Forest algorithm, which assigns an anomaly score to each indicator:
\begin{equation}
  \mathcal{I} = \left\{ e \in \mathcal{M}\,:\,
  \texttt{iforest}(\eta_e,c) = \texttt{outlier} \right\}
\end{equation}
\begin{center}
  \begin{minipage}{.9\textwidth}
    \centering
    \textit{``Mark elements $e$ whose error indicator $\eta_e$ is detected as an outlier by the Isolation Forest algorithm with a contamination $c\in[0,1]$.''}
  \end{minipage}
\end{center}
Elements are marked for refinement if their indicator is classified as an outlier. 
As the method detects all outliers, including those of small magnitude, we restrict the selection to outliers whose error indicators exceed the mean value.
As these outliers are marked and refined, the mesh might approach an approximately equidistributed error distribution. 
The parameter $c\!\in\![0,1]$, referred to as the contamination. It specifies the fraction of outliers compared to the dataset. 
Larger values of $c$ lead to more elements being marked, while smaller values restrict the selection to only the most isolated elements. 
Alternatively, an absolute threshold of $0.5$ for the anomaly score with the range $(0,1]$ can be set, instead of defining $c$ as a fraction. 
In this case, elements with an anomaly score above $0.5$ are marked as outliers.
We denote this as $c\!=\!\texttt{auto}$, since it is a non-tunable parameter on the normalized score of the algorithm.
The Isolation Forest has linear time complexity and low memory requirements \cite{liu2008isolation}, making it preferable to other outlier detection methods, such as Local Outlier Factor, which have higher time complexity \cite{cheng2019outlier}. 
The method relies on the assumption that outliers are few and sufficiently distinct from the bulk of the data. Therefore, its effectiveness may depend on the distribution of the error indicators. 
In this study, the Isolation Forest implementation from \texttt{scikit-learn} is used \cite{pedregosa2011scikit}.
\section{Benchmark problems}
\label{sec:benchmarks}
To enable a comprehensive assessment of the marking strategies, we consider established benchmark problems from solid and fluid mechanics. 
For each class, we select one problem which exact solution is sufficiently regular and one that exhibits reduced regularity, such as corner singularities. 
It is well known that the asymptotic convergence behavior of finite element approximations\textemdash{}and thus the effect of mesh refinement and marking strategies\textemdash{}depends crucially on the regularity of the exact solution\cite{ainsworth1997posteriori,dorfler1996convergent,ciarlet2002finite,grisvard2011elliptic}. 
All problems are solved with our in-house multi-physics finite element solver \texttt{XNS}.
\begin{table}
  \centering
  \begin{tabular}{c|c|c|c|c|c|c|c}
    \toprule
    ~Section~ & ~Case ID~ & ~Solid~ & ~Fluid~ & ~Singularity~ & ~Element~ & ~Ansatz~ & ~Stabilization~\\
    \midrule
    \ref{sssec:slitpanel}          & ~\texttt{SC0}~ & $\checkmark$ & $\times$ & $\times$     & Triangle & \texttt{P1}   & $\times$ \\
    \ref{sssec:infplatewithhole}   & ~\texttt{SC1}~ & $\checkmark$ & $\times$ & $\checkmark$ & Triangle & \texttt{P1}   & $\times$ \\
    \midrule
    \ref{sssec:liddrivencavity}    & ~\texttt{FC0}~ & $\times$ & $\checkmark$ & $\times$     & Triangle & \texttt{P1P1} & GLS \\
    \ref{sssec:flowaroundcylinder} & ~\texttt{FC1}~ & $\times$ & $\checkmark$ & $\checkmark$ & Triangle & \texttt{P1P1} & GLS \\
    \bottomrule
  \end{tabular}
  \label{tab:benchmarks}
  \caption{Overview of the benchmark problems. \textit{Legend}: $\checkmark$ (yes), $\times$ (no), \texttt{P1} (1\textsuperscript{st} order Lagrange Ansatz), \texttt{P1P1} (\texttt{P1} for velocity and pressure), GLS (Galerkin-Least-Squares stabilization \cite{donea2003finite}).} 
\end{table}
\subsection{Solid mechanics benchmark problems: \texttt{SC0} and \texttt{SC1}}
\label{ssec:solidbenchmarks}
The displacement field $\bs{u}(\bs{x})$ of a body occupying the domain $\bs{x}\in\Omega \subset \mathbb{R}^d$, with the spatial dimension $d$, under stationary conditions and linear elastic material behavior satisfies:
\begin{subequations}
  \begin{align}
    -\nabla\cdot\bs{\sigma} &= \bs{b} &&\text{in } \Omega \\
    \bs{\sigma}(\bs{\varepsilon}) &= \mathbb{C}(E,\nu) : \bs{\varepsilon} &&\text{in } \Omega \\
    \bs{\varepsilon}(\bs{u}) &= \tfrac{1}{2}\bigl(\nabla\bs{u} + (\nabla\bs{u})^T\bigr) &&\text{in } \Omega \\
    \bs{u} &= \bs{u}_D &&\text{on } \Gamma_D \\
    \bs{\sigma}\cdot\bs{n} &= \bs{t}_N &&\text{on } \Gamma_N
  \end{align}
  \label{eq:solidpde}
\end{subequations}
with the Cauchy stress $\bs{\sigma}$, body force $\bs{b}$, isotropic linear elasticity tensor $\mathbb{C}$, linear strain $\bs{\varepsilon}$ and prescribed Dirichlet $\bs{u}_D$ and Neumann $\bs{t}_N$ boundary conditions on $\Gamma_D\cap\Gamma_N=\emptyset \land \Gamma_D\cup\Gamma_N = \partial\Omega$ with surface normal $\bs{n}$.
\begin{figure}[!ht]
  \centering
  \begin{minipage}[t]{.45\textwidth}
    \centering
    \includegraphics[width=\textwidth]{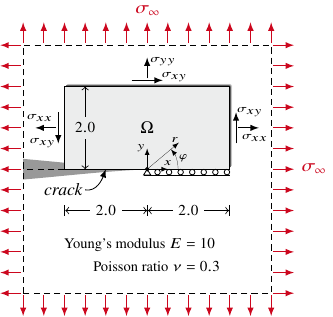}
    \caption{The plane strain slit panel problem (\texttt{SC0}) \cite{szabo1986}: pure mode I loading of a \textit{crack}.}
    \label{fig:slit-panel-setup}
  \end{minipage}
  \hfill
  \begin{minipage}[t]{.45\textwidth}
    \centering
    \includegraphics[width=\textwidth]{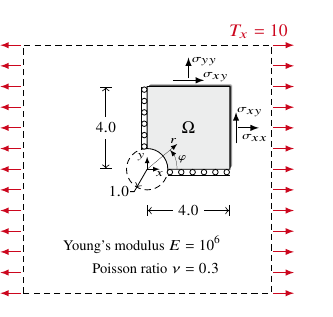}
    \caption{The plane stress infinite plate with a hole problem (\texttt{SC1}) \cite{sadd2009elasticity}: square cut-out under tension.}
    \label{fig:inf-plate-w-hole-setup}
  \end{minipage}
\end{figure}
\subsubsection {Slit panel problem (\texttt{SC0})}
\label{sssec:slitpanel}
The exact stress tensor $\bs{\sigma}$ with components $[\bs{\sigma}]_{11} = \sigma_{xx}$, $[\bs{\sigma}]_{12} = [\bs{\sigma}]_{21} = \sigma_{xy}$ and $[\bs{\sigma}]_{22} = \sigma_{yy}$ is analytically derived in polar coordinates $(r,\varphi)$ and is used as Neumann boundary conditions ($\bs{\sigma} \cdot \bs{n}$) on the loaded edges \cite{szabo1986} (see Fig. \ref{fig:slit-panel-setup}) with
\begin{subequations} \label{eq:slitpanelexactstresses}
  \begin{align}
    \sigma_{xx}(r,\varphi) &=
      \tfrac{1}{\sqrt{2\pi r}}\cos\tfrac{\varphi}{2}\bigl(1 - \sin\tfrac{\varphi}{2}\sin\tfrac{3\varphi}{2}\bigr)\,, \\
    \sigma_{yy}(r,\varphi) &=
      \tfrac{1}{\sqrt{2\pi r}}\cos\tfrac{\varphi}{2}\bigl(1 + \sin\tfrac{\varphi}{2}\sin\tfrac{3\varphi}{2}\bigr)\,, \\
    \sigma_{xy}(r,\varphi) &=
      \tfrac{1}{\sqrt{2\pi r}}\sin\tfrac{\varphi}{2}\cos\tfrac{\varphi}{2}\cos\tfrac{3\varphi}{2}\,.
  \end{align}
\end{subequations}
From this, the exact strain energy can be obtained by
\begin{equation} \label{eq:strainEnergySlitPanel}
  \mathcal{U}(\bs{u})
  = \frac{1}{2}
  \int_\Omega\bs{\sigma}(\bs{\varepsilon}):\underbrace{\mathbb{C}^{-1}:\bs{\sigma}(\bs{\varepsilon})}_{\bs{\varepsilon}(\bs{u})}\,d\Omega = 0.04741292416402
\end{equation}
where $\mathbb{C}^{-1}$ is the \textit{plane strain} compliance tensor (in matrix representation, the elasticity matrix inverse) of Hooke's law $\bs{\sigma}=\mathbb{C}:\bs{\varepsilon}$ considering the linear strain $\bs{\varepsilon}(\bs{u})$.
From \eqref{eq:slitpanelexactstresses} we can already see that a singularity in all stress components occurs at the crack tip where $r=0$ which makes it a particularly interesting case for mesh refinement to resolve this highly localized spike to infinity.
\subsubsection{Infinite plate with hole problem (\texttt{SC1})}
\label{sssec:infplatewithhole}
A second, similarly well-known problem \cite{sadd2009elasticity,cottrell2009isogeometric,schillinger2015finite} as the first one, is investigated.
Similar to the first benchmark in \ref{sssec:slitpanel}, analytically derived stresses \cite{sadd2009elasticity} are applied as Neumann boundary conditions, as shown in Fig. \ref{fig:inf-plate-w-hole-setup}, with
\begin{subequations}
  \begin{align}
   \sigma_{xx}(r,\varphi) &= \tfrac{5 \left(2 r^{4} - 16 r^{2} \sin^{4}{\mspace{-4mu}\varphi}+ 22 r^{2} \sin^{2}{\mspace{-4mu}\varphi}- 5 r^{2} + 24 \sin^{4}{\mspace{-4mu}\varphi}- 24 \sin^{2}{\mspace{-4mu}\varphi}+ 3\right)}{r^{4}} \\
   \sigma_{yy}(r,\varphi) &= \tfrac{5 \left(16 r^{2} \sin^{4}{\mspace{-4mu}\varphi}- 14 r^{2} \sin^{2}{\mspace{-4mu}\varphi}+ r^{2} - 24 \sin^{4}{\mspace{-4mu}\varphi}+ 24 \sin^{2}{\mspace{-4mu}\varphi}- 3\right)}{r^{4}} \\
   \sigma_{xy}(r,\varphi) &= \tfrac{10 \left(8 r^{2} \sin^{2}{\mspace{-4mu}\varphi}- 5 r^{2} - 12 \sin^{2}{\mspace{-4mu}\varphi}+ 6\right) \sin{\mspace{-4mu}\varphi}\cos{\varphi}}{r^{4}}
  \end{align}
\end{subequations}
which results in the exact strain energy value
\begin{equation}
  \mathcal{U}(\bs{u}) = 0.008444912711942
\end{equation}
where $\mathbb{C}^{-1}$ in \eqref{eq:strainEnergySlitPanel} is now the \textit{plane stress} compliance tensor. Here, the solution is a regular field.
\subsubsection{Convergence metric for the solid mechanics problems}
\label{sssec:convergencemetricsolid}
For the solid mechanics problems, the exact strain energy is available for both benchmark problems. 
Hence, the relative error in energy norm is calculated as
\begin{equation}
  \lVert e \rVert_\mathcal{U} = \sqrt{\tfrac{\lvert\mathcal{U}(\bs{u}) - \mathcal{U}(\bs{u}^h)\rvert}{\lvert\mathcal{U}(\bs{u})\rvert}}
\end{equation}
where $\bs{u}^h$ denotes the finite element displacement solution and $\bs{u}$ the exact solution.
\subsection{Fluid mechanics benchmark problems: \texttt{FC0} and \texttt{FC1}}
\label{ssec:fluidbenchmarks}
Assuming a steady incompressible flow in a domain $\bs{x}\in\Omega\subset\mathbb{R}^d$, the velocity field $\bs{u}(\bs{x})$ and pressure field $p(\bs{x})$ satisfy the Navier-Stokes equations:
\begin{subequations}
  \begin{align}
    -\mu\Delta\bs{u} + (\bs{u}\cdot\nabla) \bs{u} + \nabla p &= \bs{f} &&\text{in }\Omega\\
    \nabla\cdot\bs{u} &= 0 &&\text{in }\Omega\\
    \bs{u} &= \bs{u}_D &&\text{on } \Gamma_D \\
    (\mu\nabla\bs{u}-p\bs{I})\cdot\bs{n} &= \bs{t}_N &&\text{on } \Gamma_N
  \end{align}
\end{subequations}
with the dynamic viscosity $\mu$, while the remaining parts of the PDE are the velocity equivalent of the PDE \eqref{eq:solidpde} in \ref{ssec:solidbenchmarks}.
\begin{figure}[!ht]
  \centering
  \begin{minipage}[t]{.35\textwidth}
    \centering
    \includegraphics[width=.95\textwidth]{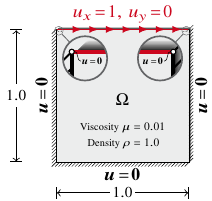}
    \caption{The lid-driven cavity problem (\texttt{FC0}) \cite{ghia1982high}: Fluid in a cavity with moving top wall.}
    \label{fig:lid-driven-cavity-setup}
  \end{minipage}
  \hfill
  \begin{minipage}[t]{.6\textwidth}
    \centering
    \includegraphics[width=\textwidth]{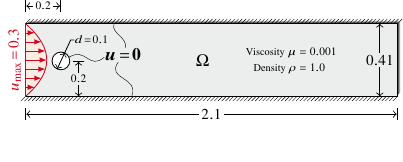}
    \caption{The flow around cylinder problem (\texttt{FC1}) \cite{schafer1996benchmark}: Steady flow in laminar regime with $\mathit{Re}=20$.}
    \label{fig:flow-around-cylinder-setup}
  \end{minipage}
\end{figure}
\subsubsection{Lid-driven cavity problem (\texttt{FC0})}
\label{sssec:liddrivencavity}
The first fluid mechanics benchmark case is the well-known flow in a cavity with a moving lid \cite{ghia1982high} as depicted in Fig. \ref{fig:lid-driven-cavity-setup}.
Since the top boundary moves with a constant velocity and the lateral boundaries are non-slip, two configurations are possible: (i) the corner nodes belongs to the moving boundary (\textit{leaky}) or (ii) they belong to the non-slip boundaries left and right (\textit{non-leaky}).
Here, we choose the \textit{non-leaky} boundary conditions (see Fig. \ref{fig:lid-driven-cavity-setup}) which provoke pressure singularities in the top left and right corner.
\subsubsection{Flow around cylinder problem (\texttt{FC1})}
\label{sssec:flowaroundcylinder}
The second classical fluid mechanics benchmark is the flow around a cylindrical object \cite{schafer1996benchmark} shown in Fig. \ref{fig:flow-around-cylinder-setup}. 
The chosen set of parameters result in the Reynolds number $\mathit{Re} = 2\rho u_\mathit{max}d/3\mu=20$.
While the boundary layers at the non-slip boundaries generally invoke high velocity gradients normal to the surface, no irregular features like singularities are present. 
\subsubsection{Convergence metric for the fluid mechanics problems}
\label{sssec:convergencemetricfluid}
From a so-called \textit{overkill} solution on a mesh with $9\,241\,134$ elements and $4\,624\,568$ nodes for \texttt{FC0}, and $7\,908\,196$ elements and $3\,959\,434$ nodes for \texttt{FC1}, we obtain the velocity $\bs{u}^\star$.
We can then compute the $L^2$-norm error of the adaptively refined velocity solution $\bs{u}^h$ with
\begin{equation}
  \lVert e \rVert_{L^2} = 
  \tfrac{\lVert \bs{u}^h - \bs{u}^\star \rVert_{L^2(\Omega)}}{\lVert\bs{u}^\star\rVert_{L^2(\Omega)}} = 
  \sqrt{ \tfrac{\int_\Omega \lvert\bs{u}^h - \bs{u}^\star \rvert^2\,d\Omega}{\int_\Omega \lvert\bs{u}^\star \rvert^2\,d\Omega} }.
\end{equation}
\section{Results}
\label{sec:results}
A marking strategy can be effective and efficient in two ways: 
  (i) it can minimize the number of refinement cycles needed to reach a sufficiently low error, and 
  (ii) it can increase the convergence rate to obtain a better accuracy for a given number of nodes or elements. 
It should be noted that (i) can be cheaper in total, despite the worse convergence rate while (ii) seems more optimal for the individual meshes in a sense that the refinement marking is exceptionally precise in choosing the elements for refinement.
The total cost of an adaptive simulation is the sum of the meshing and solution times over all cycles, $\sum_i \bigl(T_\mathrm{mesh}(\mathcal{M}_i) + T_\mathrm{solve}(\mathcal{M}_i)\bigr)$, and thus depends both on the number of cycles and on the size of the meshes $\mathcal{M}_i$ visited along the way.
Each cycle carries a fixed overhead (mesh generation, assembly, and solver setup), so strategies that need fewer cycles are advantageous even when their final mesh is slightly larger, whereas strategies that refine too aggressively reach the target error in few cycles but on an unnecessarily fine mesh.
The cost of the marking step itself is limited to sorting or thresholding of the element indicators for \texttt{DOE}, \texttt{QUA}, \texttt{MAX}, and \texttt{ZSC}, and \texttt{ISO} scales linearly with the number of elements \cite{liu2008isolation}.
For \texttt{MAX}, \texttt{QUA}, \texttt{DOE}, \texttt{ZSC}, marking accounted for less than 0.1\% of the time per cycle in all cases, whereas for \texttt{ISO}, it was typically on the order of 1\% and could reach up to 10\% in early cycles.
Hence, we focus on how the marking strategy affects the overall efficiency of the adaptive simulation rather than on the efficiency of the marking step itself.
The refinement is conducted by reducing the element size of the marked elements by a factor of $0.5$. 
The value of $0.5$ is chosen to align with standard element subdivision, which ensures a consistent refinement procedure and isolates the effect of the marking strategy on the overall AMR performance.
The best parameters in the tested ranges and the corresponding required refinement cycles to converge for each strategy are summarized in Tab. \ref{tab:bestmarkingparameters}.
For every case, a relative error of $1\%$ is chosen as the criterion to consider the solution as converged.
\texttt{QUA} and \texttt{ZSC} are the most robust strategies, with good performance in both the solid and fluid mechanics problems. 
\texttt{DOE} and \texttt{MAX} vary more strongly: they perform poorly in \texttt{SC0} and \texttt{FC0}, but competitively in \texttt{SC1} and \texttt{FC1}.
\texttt{ISO} performs competitively in the fluid but worse in the solid mechanics problems.
Among all marking strategies, the best-performing parameter is the same for all benchmark problems, with only very few exceptions.
\begin{table}
    \centering
    \setlength{\tabcolsep}{3.5pt}   
    \small                        
    \begin{tabular}{c
        |ccr
        |ccr
        |ccr
        |ccr
        |ccr}
        \toprule
        &
        \multicolumn{3}{c|}{\texttt{DOE}} &
        \multicolumn{3}{c|}{\texttt{ZSC}} &
        \multicolumn{3}{c|}{\texttt{MAX}} &
        \multicolumn{3}{c|}{\texttt{QUA}} &
        \multicolumn{3}{c}{\texttt{ISO}} \\
        \texttt{ID} & $\theta$ & $i$ & \multicolumn{1}{c|}{$n_n$} &
          $z^*$ & $i$ & \multicolumn{1}{c|}{$n_n$} &
          $\alpha$ & $i$ & \multicolumn{1}{c|}{$n_n$} &
          $\gamma$ & $i$ & \multicolumn{1}{c|}{$n_n$} &
          $c$ & $i$ & \multicolumn{1}{c}{$n_n$} \\ \midrule
        \texttt{SC0} &
        0.9  & \cellcolor{red-10}\texttt{\color{red}19} & 99.8k &
        0.25 & \cellcolor{green-10}\texttt{\color{green}12} & 98.9k &
        -  & \cellcolor{red-10}\texttt{\color{red}-} & - &
        0.6  & \cellcolor{green-10}\texttt{\color{green}11} & 216.9k &
        0.5  & \cellcolor{orange-10}\texttt{\color{orange}14} & 72.1k  \\
        \texttt{SC1} &
        0.9  & \cellcolor{green-10}\texttt{\color{green}7} & 6.2k &
        0.25 & \cellcolor{orange-10}\texttt{\color{orange}9} & 5.9k &
        0.1  & \cellcolor{orange-10}\texttt{\color{orange}8} & 6.0k &
        0.1  & \cellcolor{green-10}\texttt{\color{green}6} & 10.7k &
        0.4  & \cellcolor{red-10}\texttt{\color{red}20} & 5.9k  \\
        \texttt{FC0} &
        0.9  & \cellcolor{red-10}\texttt{\color{red}34} & 10.0k &
        0.25 & \cellcolor{orange-10}\texttt{\color{orange}10} & 36.8k &
        0.1  & \cellcolor{red-10}\texttt{\color{red}33} & 8.1k &
        0.1  & \cellcolor{green-10}\texttt{\color{green}5} & 27.7k &
        0.5  & \cellcolor{orange-10}\texttt{\color{orange}9} & 30.1k \\
        \texttt{FC1} &
        0.9  & \cellcolor{green-10}\texttt{\color{green}5} & 4.5k &
        0.25 & \cellcolor{green-10}\texttt{\color{green}5} & 3.6k &
        0.1  & \cellcolor{orange-10}\texttt{\color{orange}7} & 4.9k &
        0.1  & \cellcolor{green-10}\texttt{\color{green}5} & 30.4k &
        0.5  & \cellcolor{red-10}\texttt{\color{red}8} & 5.0k \\
        \bottomrule
    \end{tabular}
    \caption{Best parameters w.r.t. minimal number of refinement cycles $i$ and number of nodes $n_n$ until the error threshold is reached for each case and marking strategy. Here, \emph{cyc.} denotes the number of refinement cycles and $n_n$ the number of nodes.}
    \label{tab:bestmarkingparameters}
\end{table}
\subsection{Solid mechanics benchmark results (\texttt{SC0} and \texttt{SC1})}
\begin{figure}
  \begin{minipage}{\textwidth}
    \centering
    \includegraphics[width=.95\textwidth]{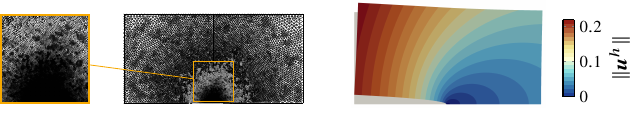}
    \vspace*{-1mm}
    \caption{\texttt{SC0} \textendash{} Refined mesh and displacement field in deformed configuration (\texttt{DOE}, $\theta=0.9$).}
    \label{fig:slit-panel-images}
  \end{minipage}\\[2mm]
  \begin{minipage}{\textwidth}
    \centering
    \includegraphics[width=\textwidth]{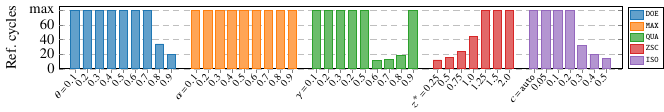}
    \vspace*{-5mm}
    \caption{\texttt{SC0} \textendash{} Required refinement cycles per marking strategy until error $\lVert e \rVert_\mathcal{U}<1\%$.}
    \label{fig:slit-panel-convergence-stats}
  \end{minipage}\\[2mm]
  \begin{minipage}{\textwidth}
    \centering 
    \begin{minipage}[t]{.42\textwidth}
      \centering
      \includegraphics[width=.96\textwidth]{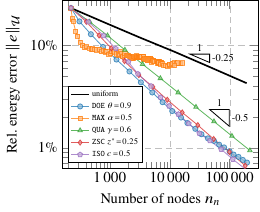}
      \caption{\texttt{SC0} \textendash{} Strain energy convergence with parameters of min. refinement cycles.}
      \label{fig:slit-panel-error-convergence}
    \end{minipage}
    \hfill
    \begin{minipage}[t]{.56\textwidth}
      \centering
      \includegraphics[width=\textwidth]{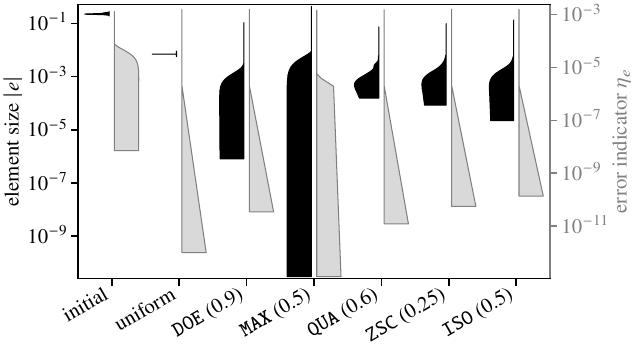}
      \caption{\texttt{SC0} \textendash{} Distribution of element size and error indicator in the final meshes.}
      \label{fig:slit-panel-violin}
    \end{minipage}
  \end{minipage}
\end{figure}
\begin{figure}
  \begin{minipage}{\textwidth}
    \centering 
    \includegraphics[width=.8\textwidth]{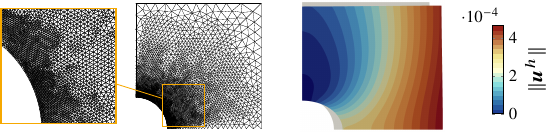}
    \vspace*{-1mm}
    \caption{\texttt{SC1} \textendash{} Refined mesh and displacement field in deformed configuration (\texttt{DOE}, $\theta=0.9$).}
    \label{fig:inf-plate-w-hole-images}
  \end{minipage}\\[2mm]
  \begin{minipage}{\textwidth}
    \centering
    \includegraphics[width=\textwidth]{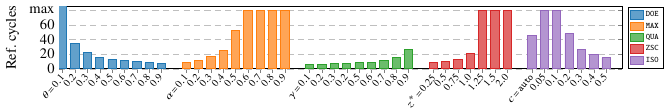}
    \vspace*{-5mm}
    \caption{\texttt{SC1} \textendash{} Required refinement cycles per marking strategy until error $\lVert e \rVert_\mathcal{U}<1\%$.}
    \label{fig:inf-plate-w-hole-convergence-stats}
  \end{minipage}\\[2mm]
  \begin{minipage}{\textwidth}
    \centering 
    \begin{minipage}[t]{.42\textwidth}
      \centering
      \includegraphics[width=.96\textwidth]{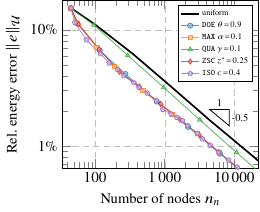}
      \caption{\texttt{SC1} \textendash{} Strain energy convergence with parameters of min. refinement cycles.}
      \label{fig:inf-plate-w-hole-error-convergence}
    \end{minipage}
    \hfill
    \begin{minipage}[t]{.56\textwidth}
      \centering
      \includegraphics[width=\textwidth]{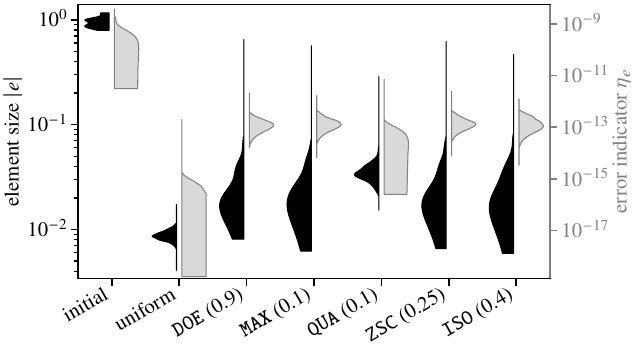}
      \caption{\texttt{SC1} \textendash{} Distribution of element size and error indicator in the final meshes.}
      \label{fig:inf-plate-w-hole-violin}
    \end{minipage}
  \end{minipage}
\end{figure}
Figures \ref{fig:slit-panel-images} and \ref{fig:inf-plate-w-hole-images} show the refined mesh and displacement field. 
In \texttt{SC0}, the adaptive strategies concentrate refinement at the stress singularity, while in \texttt{SC1} the refinement follows the more regular displacement field with smaller elements around the circular hole.
Figures \ref{fig:slit-panel-error-convergence} and \ref{fig:inf-plate-w-hole-error-convergence} show the error convergence of the strain energy. 
In \texttt{SC0}, all adaptive strategies except \texttt{MAX} reach an error below $1\%$ at about $10^5$ nodes, while uniform refinement converges with about half of the optimal rate for a linear finite element ansatz. 
In \texttt{SC1}, all strategies, including uniform refinement, show nearly identical and optimal convergence rates, which reflects the regular solution field.
Figures \ref{fig:slit-panel-convergence-stats} and \ref{fig:inf-plate-w-hole-convergence-stats} show the required refinement cycles to converge for each marking parameter. 
\texttt{DOE} performs well for $\theta \geq 0.8$ in both \texttt{SC0} and \texttt{SC1}, but fails to converge for $\theta < 0.8$ in \texttt{SC0}. 
\texttt{MAX} has a high initial convergence in \texttt{SC0}, but then stagnates.
In \texttt{SC1} it works well for $\alpha = 0.1$\textendash{}$0.3$. 
\texttt{QUA} is efficient for $\gamma = 0.6,\dots,0.8$ in \texttt{SC0}, while all $\gamma$ values work well in \texttt{SC1}. 
Values of $\gamma < 0.6$ lead to over-refinement in \texttt{SC0} and exceed the cycle limit. 
\texttt{ZSC} performs well for $z^* \leq 1$ in both problems, but better in \texttt{SC1}. 
\texttt{ISO} fails for $c \leq 0.1$, while higher values give average performance.
Figures \ref{fig:slit-panel-violin} and \ref{fig:inf-plate-w-hole-violin} show the violin plots of element size and error indicator after the final cycle. 
In \texttt{SC0}, \texttt{QUA}, \texttt{ZSC}, and \texttt{ISO} produce narrow and concentrated element size distributions, with \texttt{QUA} being the most concentrated. 
\texttt{MAX} shows a broad distribution with a long tail toward smaller elements, indicating over-refinement. 
In \texttt{SC0}, \texttt{QUA}, \texttt{ZSC} and \texttt{ISO} also achieve many low-magnitude error indicators and only very few larger ones, which mainly occur at the singularity. 
\texttt{DOE} gives a less concentrated element size distribution but still low error indicators, while \texttt{MAX} produces a wider spread of higher error indicators and thus weaker error suppression. 
In \texttt{SC1}, the element sizes are more spread out and closer to a normal distribution than in \texttt{SC0} because of the more regular solution field. 
The error indicator distributions are also more compact and close to normal for almost all strategies. 
\texttt{QUA} shows a wider spread in the error indicators, closer to uniform refinement, which follows from the low $\gamma$ value of 0.1.
\subsection{Fluid mechanics benchmark results  (\texttt{FC0} and \texttt{FC1})}
\def\fluidErrorMetric{rel_L2_velocity}
\begin{figure}
  \begin{minipage}{\textwidth}
    \centering 
    \includegraphics[width=.8\textwidth]{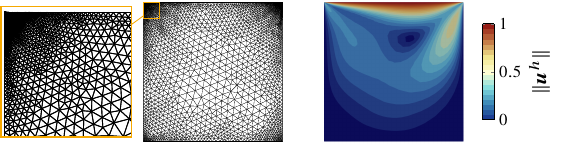}
    \vspace*{-1mm}
    \caption{\texttt{FC0} \textendash{} Refined mesh and velocity field (\texttt{DOE}, $\theta=0.9$).}
    \label{fig:lid-driven-cavity-images}
  \end{minipage}\\[2mm]
  \begin{minipage}{\textwidth}
    \centering
    \includegraphics[width=.8\textwidth]{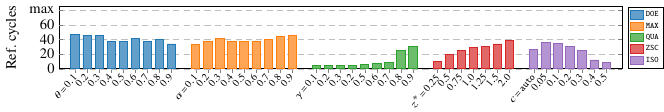}
    \vspace*{-5mm}
    \caption{\texttt{FC0} \textendash{} Required refinement cycles per marking strategy until $\lVert e \rVert_{L^2}<1\%$.}
    \label{fig:lid-driven-cavity-convergence-stats}
  \end{minipage}\\[2mm]
  \begin{minipage}{\textwidth}
    \centering 
    \begin{minipage}[t]{.42\textwidth}
      \centering
      \includegraphics[width=.96\textwidth]{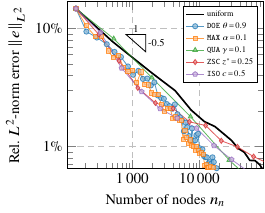}
      \caption{\texttt{FC0} \textendash{} $L^2$ error convergence with parameters of min. refinement cycles.}
      \label{fig:lid-driven-cavity-error-convergence}
    \end{minipage}
    \hfill
    \begin{minipage}[t]{.56\textwidth}
      \centering
      \includegraphics[width=\textwidth]{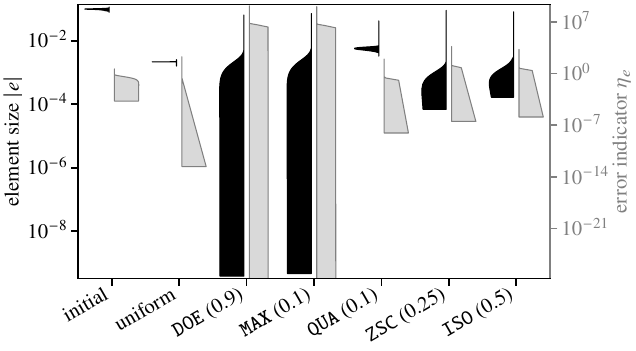}
      \caption{\texttt{FC0} \textendash{} Distribution of element size and error indicator in the final meshes.}
      \label{fig:lid-driven-cavity-violin}
    \end{minipage}
  \end{minipage}
\end{figure}
\begin{figure}
  \begin{minipage}{\textwidth}
    \centering
    \includegraphics[width=.96\textwidth]{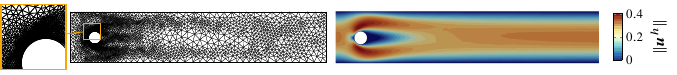}
    \vspace*{-2mm}
    \caption{\texttt{FC1} \textendash{} Refined mesh and velocity field (\texttt{DOE}, $\theta=0.9$).}
    \label{fig:flow-around-cylinder-images}
  \end{minipage}\\[2mm]
  \begin{minipage}{\textwidth}
    \centering
    \includegraphics[width=\textwidth]{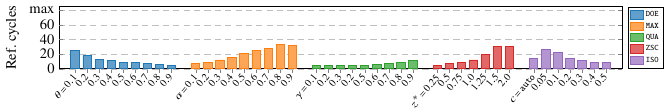}
    \vspace*{-5mm}
    \caption{\texttt{FC1} \textendash{} Required refinement cycles per marking strategy until error $\lVert e \rVert_{L^2}<1\%$.}
    \label{fig:flow-around-cylinder-convergence-stats}
  \end{minipage}\\[2mm]
  \begin{minipage}{\textwidth}
    \centering 
    \begin{minipage}[t]{.42\textwidth}
      \centering
      \includegraphics[width=.96\textwidth]{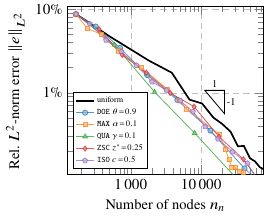}
      \caption{\texttt{FC1} \textendash{} $L^2$ error convergence with parameters of min. refinement cycles.}
      \label{fig:flow-around-cylinder-error-convergence}
    \end{minipage}
    \hfill
    \begin{minipage}[t]{.56\textwidth}
      \centering
      \includegraphics[width=\textwidth]{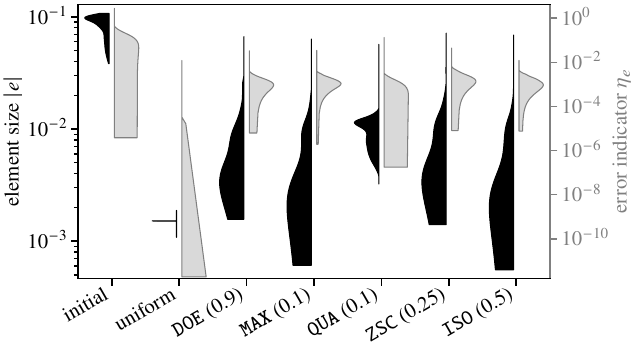}
      \caption{\texttt{FC1} \textendash{} Distribution of element size and error indicator in the final meshes.}
      \label{fig:flow-around-cylinder-violin}
    \end{minipage}
  \end{minipage}
\end{figure}
Figures \ref{fig:lid-driven-cavity-images} and \ref{fig:flow-around-cylinder-images} show the refined mesh and velocity field. 
In \texttt{FC0}, refinement is focused on the top left and right corners because of the strong local gradients caused by the abrupt boundary condition changes. 
In \texttt{FC1}, refinement appears around the cylinder and in the wake region, where the velocity gradients are highest.
Figures \ref{fig:lid-driven-cavity-error-convergence} and \ref{fig:flow-around-cylinder-error-convergence} show the $L^2$-norm error convergence. 
In \texttt{FC0}, all adaptive strategies reach an error below $1\%$ faster than uniform refinement, but not by a large margin. 
The convergence rates are less consistent during the refinement process and are on average similar to uniform refinement, since the $L^2$-norm does not capture gradients. 
In \texttt{FC1}, the convergence rates are more consistent than in \texttt{FC0}, but the average behavior is still similar to uniform refinement.
Figures \ref{fig:lid-driven-cavity-convergence-stats} and \ref{fig:flow-around-cylinder-convergence-stats} show the required refinement cycles to converge for each marking parameter. 
\texttt{DOE} performs only moderately in \texttt{FC0} for all $\theta$ values, but very well in \texttt{FC1}, with the best result at $\theta = 0.9$. 
\texttt{MAX} is also only moderate in \texttt{FC0}, but good in \texttt{FC1}, with the best result at $\alpha = 0.1$. 
\texttt{QUA} is very efficient for low $\gamma$ values around $0.1$, while all $\gamma$ values in \texttt{FC1} perform similarly well. 
In \texttt{FC0}, $\gamma \geq 0.8$ causes under-refinement and exceeds the cycle limit for $\gamma = 0.8$. 
\texttt{ZSC} works well for low $z^*$ values around $0.25$ in both problems, but better in \texttt{FC1}.
\texttt{ISO} behaves similarly, with good results for higher $c$ values and performance close to \texttt{ZSC}.
Figures \ref{fig:lid-driven-cavity-violin} and \ref{fig:flow-around-cylinder-violin} show the violin plots of element size and error indicator after the final iteration. 
In \texttt{FC0}, \texttt{QUA} gives extremely narrow and concentrated element size distributions, similar to uniform refinement, while \texttt{ZSC} and \texttt{ISO} are less concentrated and contain more small elements. 
\texttt{DOE} and \texttt{MAX} show broader distributions with a tail toward smaller elements. 
In \texttt{FC0}, \texttt{QUA}, \texttt{ZSC}, and \texttt{ISO} again produce many low-magnitude error indicators and only very few larger ones, while \texttt{DOE} and \texttt{MAX} produce a wider spread of higher error indicators. 
In \texttt{FC1}, the element size distribution is more spread than in \texttt{FC0} because of the more regular solution field, and \texttt{QUA} has the most compact distribution. 
\texttt{DOE}, \texttt{MAX}, \texttt{ZSC}, and \texttt{ISO} yield quite concentrated, normally distributed error indicators, while \texttt{QUA} shows more features of a uniform error distribution.
\subsection{Limitations and transferability}
\label{ssec:limitations}
The results are based on four two-dimensional, stationary benchmarks, linear elements, and the Kelly estimator.
The ranking of the strategies should therefore be read as a hypothesis for other settings rather than a general result.
We nevertheless expect some tendencies to persist: strategies acting on the distribution of the indicators (\texttt{QUA}, \texttt{ZSC}, \texttt{ISO}) are independent of their absolute magnitude, whereas \texttt{MAX} depends on a single extreme value and should remain fragile wherever isolated spikes occur, e.g., at singularities or localized nonlinear effects.
Other aspects may change: in 3D, refining a marked element adds more elements than in 2D, so \texttt{QUA} may require a larger $\gamma$ to avoid over-refinement, and in transient or nonlinear problems the error distribution evolves, so parameters may need re-tuning and coarsening becomes relevant.
Testing these expectations is the aim of the future work outlined in \ref{sec:conclusion}.
\section{Conclusion}
\label{sec:conclusion}
Our work highlights the following key findings for each of the marking strategies:
\begin{itemize}
  \item \texttt{DOE} performs robustly for $\theta \geq 0.8$ and offers a good balance between the number of refinement cycles and the convergence rate, but depends strongly on the problem. 
  \item \texttt{MAX} can perform well for low $\alpha$ values, but is sensitive when the solution field has reduced regularity, especially if singularities are captured by the error estimator through the solution gradient. 
  In the fluid mechanics problems, \texttt{MAX} behaves in a less sensitive way. 
  \item \texttt{QUA} reaches convergence with few refinement cycles and produces predictable mesh sizes because it always refines the same fraction of the mesh. 
  A low $\gamma \approx 0.1$ required the fewest cycles in \texttt{SC1}, \texttt{FC0}, and \texttt{FC1}, but led to over-refinement in \texttt{SC0}, where $\gamma \approx 0.6$ was needed; $\gamma \geq 0.8$ causes under-refinement. 
  By allowing a few more refinement cycles due to larger $\gamma$, the convergence rate can be improved to nearly optimal. 
  \item \texttt{ZSC} is a good alternative to \texttt{DOE} because it balances the number of refinement cycles and the convergence rate well. 
  The value $z^* \approx 0.25$ was optimal in all four benchmarks, and in \texttt{FC0}, \texttt{ZSC} reduced the number of refinement cycles compared to \texttt{DOE} while giving similar results. 
  \item \texttt{ISO} performs well for large contamination values ($c \approx 0.4$--$0.5$) and is comparable to \texttt{ZSC}, but usually needs slightly more refinement cycles. 
\end{itemize}
\textbf{Practical recommendation.}
For stationary problems of the type studied here and the Kelly error estimator, we suggest \texttt{ZSC} marking with $z^* \approx 0.25$, applied to logarithmically scaled error indicators, as a robust default: it reached the 1\% error threshold within 5 to 12 cycles for all four benchmarks.
\texttt{QUA} with $\gamma \approx 0.6$ is a conservative alternative, and \texttt{DOE} with $\theta \geq 0.9$ is preferable if a criterion with a theoretical foundation is desired.
\texttt{MAX} should be used with caution when singularities are present, and \texttt{ISO} only with a high contamination.
These recommendations are restricted to the tested class of problems (see the discussion of limitations in section \ref{ssec:limitations}).
\textbf{Future work.} 
Future work could involve extending the study to nonlinear materials and transient analyses involving coarsening as well in 2D and 3D. 
Additional investigations may include alternative error estimators (e.g., Zienkiewicz-Zhu, Hessian-based) and discretizations (spline-based, high-order, space-time).
\textbf{Acknowledgements}\quad 
The authors used the large language model RWTHgpt (KI:connect.nrw) for language refinement and improving manuscript clarity. 
All scientific content, interpretations, and conclusions remain the sole responsibility of the authors.
\bibliographystyle{unsrt} 
\bibliography{references}
\end{document}